\documentclass{article}
\usepackage{spconf}
\usepackage{amsmath,amssymb,amsfonts}
\usepackage{algorithm}
\usepackage{algorithmicx}
\usepackage{algpseudocode}
\usepackage{graphicx}
\usepackage{textcomp}
\usepackage{xcolor}
\usepackage{stfloats}
\usepackage{subfigure}
\usepackage[colorlinks=true, citecolor=blue, linkcolor=blue, urlcolor=blue]{hyperref}
\usepackage{cite}
\usepackage{shortcuts}

\title{Efficient Transmit Waveform Design for MIMO-OFDM DFRC Systems with 1-Bit DACs}
\name{
Chenfei Huang$^{\star}$ \qquad
Mingjie Shao$^{\star}$ \qquad
Ya-Feng Liu$^{\dagger}$
}

\address{
$^{\star}$ State Key Laboratory of Mathematical Sciences, Academy of Mathematics and Systems Science,\\
Chinese Academy of Sciences, Beijing, China\\
$^{\dagger}$ School of Mathematical Sciences, Beijing University of Posts and Telecommunications, Beijing, China\\
Email: huangchenfei@lsec.cc.ac.cn, mingjieshao@amss.ac.cn, yafengliu@bupt.edu.cn
}

\begin{document}
\ninept
\maketitle
\begin{abstract}
This paper studies efficient waveform design for MIMO-OFDM dual-functional radar-communication (DFRC) systems with 1-bit digital-to-analog converters (DACs), which remains underexplored in the literature. 
We formulate the DFRC waveform design problem by maximizing the trace of the target angular Fisher information matrix (FIM), subject to symbol-error probability constraints realized via constructive interference (CI) and a time-domain 1-bit transmit alphabet constraint. 
The resulting problem is a large-scale nonlinear integer programming problem.
We develop an inexact alternating direction method of multipliers (ADMM) algorithm with alternating time- and frequency-domain updates.
The proposed scheme enables subcarrier-wise optimization in the frequency domain and closed-form primal and dual updates in the time domain. 
In particular, the CI constraints are deliberately retained in both the time- and frequency-domain subproblems to promote the feasibility of intermediate iterates and improve the convergence behavior of ADMM.
Simulation results demonstrate promising DFRC performance.
\end{abstract}
\section{Introduction}
\label{sec:intro}

Dual-functional radar-communication (DFRC) systems use a shared transmit platform and waveform to simultaneously serve communication users and sense radar targets \cite{hassanien2019dual}. 
By sharing spectrum and hardware resources, DFRC can improve spectrum utilization and reduce hardware costs \cite{liu2018toward}.
Multiple-input multiple-output~(MIMO) technology is a key enabling technology for DFRC, as its spatial degrees of freedom can be exploited to simultaneously support multi-user communications and radar sensing.
Prior studies on MIMO DFRC have focused on designing transmit waveforms or beamformers to balance communication and sensing performance \cite{liu2022cramer, su2022secure, wu2024efficient, zou2024energy, fang2026optimal}.
Building upon the spatial degrees of freedom offered by MIMO, orthogonal frequency-division multiplexing (OFDM) further offers flexible frequency-domain waveform design and facilitates high-rate communications as well as range-Doppler sensing, making MIMO-OFDM an attractive architecture for DFRC systems \cite{liyanaarachchi2024joint, feng2025odfm, dai2026tutorial}.
Waveform design for MIMO-OFDM DFRC systems has therefore attracted increasing research interest in recent years, with studies considering  joint transmit and receive designs  under various communication and sensing metrics \cite{johnston2022mimo,li2025mimo,wei2024waveform,liao2025design}.

On the other hand, the increasing number of antennas in MIMO DFRC systems leads to increased hardware complexity and power consumption \cite{bjornson2014massive}.
Low-resolution digital-to-analog converters~(DACs), particularly 1-bit DACs, provide a power-efficient solution for massive MIMO communication systems and have recently attracted increasing attention in DFRC systems \cite{yu2022precoding,wu2025quantized,huang2026cramer,liu2024survey}.
A major challenge in extending waveform design to MIMO-OFDM DFRC systems is the need to optimize the transmit symbols across all subcarriers, resulting in a large-scale discrete optimization problem.
To the best of our knowledge, efficient waveform design for MIMO-OFDM DFRC systems with 1-bit DACs remains underexplored and is the focus of the present paper.

In this paper, we formulate the waveform design problem and develop an efficient algorithm for MIMO-OFDM DFRC systems with 1-bit DACs.
We maximize the trace of the target angular Fisher information matrix (FIM), which is equivalent to minimizing a tractable lower bound on the Cram\'er--Rao bound (CRB), subject to symbol-error probability (SEP) constraints for communications, realized via constructive interference (CI), and the 1-bit transmit alphabet constraint.
The resulting problem is nonconvex, discrete, and large-scale, making it challenging to solve.
We develop an inexact alternating direction method of multipliers (ADMM) algorithm that alternates between time-domain and frequency-domain  updates.
Such an alternating time-frequency procedure enables subcarrier decoupling in the frequency domain and facilitates closed-form primal and dual updates in the time domain, leading to a computationally efficient implementation.
In particular, we impose the  CI constraints in both the time- and frequency-domain subproblems to promote the feasibility of intermediate iterates and improve the convergence behavior of ADMM.
Numerical simulations demonstrate improved convergence behavior and DFRC performance compared with the beampattern-based benchmark.

\section{System Model and Problem Formulation}
\label{sec:system model}

We consider a MIMO-OFDM DFRC system where a base station~(BS) with $N_t$ transmit antennas and $N_r$ receive antennas simultaneously serves $K$ single-antenna communication users and senses $M$ far-field radar targets. Each transmit antenna is equipped with a pair of 1-bit DACs.

\subsection{Communication Model and Design Metric}

We adopt the standard MIMO-OFDM signal model in \cite{studer2004broadband}. Each transmit frame consists of $L$ OFDM symbols over $N_s$ subcarriers with subcarrier spacing $\Delta f$ and duration $T_{\mathrm{sym}}$.
Let $\bx_F[n,l]\in\C^{N_t}$ denote the frequency-domain transmit vector on subcarrier $n$ in  OFDM symbol~$l$. The time-domain sample is obtained via the IDFT as

\begin{equation}
\label{eq:time-domain signal}
\bx_T[q,l] \!=\!\frac{1}{\sqrt{N_s}}\!
\sum_{n=0}^{N_s-1} \!
 \bx_F[n,l] e^{j2\pi qn/N_s},~
\forall q\in \![N_s],~\forall l\in \![L],
\end{equation}
where $[N]$ denotes $\{0,1,\ldots,N-1\}$.
Define the stacked frequency- and time-domain transmit signal vectors as
\begin{equation*}
	\begin{aligned}
		\bx_F[l]=[\bx_F^\T[0,l],\bx_F^\T[1,l],\ldots,\bx_F^\T[N_s-1,l]]^\T,\\
    	\bx_T[l]=[\bx_T^\T[0,l],\bx_T^\T[1,l],\ldots,\bx_T^\T[N_s-1,l]]^\T.
	\end{aligned}
\end{equation*}
It follows from \eqref{eq:time-domain signal} that
\begin{equation}
	\label{eq:DFT relation}
	\bx_F[l]=\bG\bx_T[l],
\end{equation}
where $\bG \triangleq \bF_{N_s} \otimes \bI_{N_t}$, $\bF_{N_s}$ is the unitary $N_s\times N_s$ DFT matrix and $\otimes$ denotes the Kronecker product.
Due to the use of 1-bit DACs, the time-domain transmit signal $\bx_T$ is constrained to a finite alphabet:
\begin{equation}
\label{eq:1-bit constraint}
\bx_T[l] \in \mathcal{X}^{N_sN_t},
\quad \forall l\in[L],
\end{equation}
where
\begin{equation*}
\mathcal{X}
=
\sqrt{\frac{P}{2N_t}}
\{\pm1\pm j\},
\end{equation*}
and $P$ denotes the transmit power.

With cyclic prefix (CP) insertion at the BS and CP removal at the users, the signal received by user $k$ on subcarrier $n$ of OFDM symbol $l$ is
\begin{equation*}
y_{k,n,l}
=
\bh_{k,n}^{\Herm}\bx_F[n,l]
+
z_{k,n,l},
\end{equation*}
where $\bh_{k,n}\in\C^{N_t}$ denotes the frequency-domain channel between the BS and user $k$ on subcarrier $n$, and $z_{k,n,l}\sim\mathcal{CN}(0,\sigma_c^2)$ is additive white Gaussian noise.

We consider a symbol-error probability (SEP) metric for all communication users.
Assume that the information symbols are drawn from an $\Omega$-ary unit-modulus PSK constellation with $\Omega\geq4$.
Let $s_{k,n,l}$ denote the symbol intended for user $k$ on subcarrier $n$ of OFDM symbol $l$.
We adopt the symbol-level CI formulation \cite{masouros2015exploiting}, which pushes the noiseless received signal deep into the decision region to reduce SEP.
A sufficient condition to guarantee an SEP no greater than $\epsilon_c$ for every user on every subcarrier can be derived from \cite{shao2018multiuser, wu2025quantized}:
\begin{equation}
\label{eq:CI constraint}
\Re\left\{\bC_{n,l}\bx_F[n,l]\right\} \geq
\boldsymbol{\gamma},
~ \forall n\in[N_s],~\forall l\in[L],
\end{equation}
where
\begin{equation*}
\boldsymbol{\gamma}
=
\frac{\sigma_c\mathrm{erfc}^{-1}(\epsilon_c)}
{\cos(\pi/\Omega)}
\mathbf{1}_{2K},
\end{equation*}
$\mathrm{erfc}(\cdot)$ denotes the complementary error function,
\begin{equation*}
\begin{split}
\bC_{n,l}= &~
\widetilde{\bH}_{n,l}
\otimes
\begin{bmatrix}
\tan(\pi/\Omega)-j\\
\tan(\pi/\Omega)+j
\end{bmatrix},\\
\widetilde{\bH}_{n,l}= &~ [
s_{1,n,l}\bh_{1,n},
s_{2,n,l}\bh_{2,n},
\ldots,
s_{K,n,l}\bh_{K,n} ]^{\Herm}.
\end{split}
\end{equation*}
The above frequency-domain CI constraints can be equivalently expressed in the time domain via \eqref{eq:DFT relation} and \eqref{eq:CI constraint} as
\begin{equation}
    \label{eq:time-domain CI}
	\Re\left\{\bD_l\bx_T[l]\right\}
      \geq \bar{\boldsymbol{\gamma}}, \quad \forall l \in [L],
\end{equation}
where $\bD_l = \bC_l\bG, \bC_l=\diag(\bC_{0,l}, \bC_{1,l}, \ldots, \bC_{N_s-1,l})$, and $\bar{\boldsymbol{\gamma}}=\mathbf{1}_{N_s} \otimes \boldsymbol{\gamma}$. 

\subsection{Sensing Model and Design Metric}

We next derive the sensing metric for target angle estimation in the considered MIMO-OFDM system.
After  DFT processing, the received echo on subcarrier $n$ of OFDM symbol $l$ is given by \cite{dai2026tutorial}
\begin{equation}
\label{eq:sensing_received_signal}
\begin{aligned}
\by_s[n,l]
=&\sum_{m=1}^{M}
\alpha_m
\ba_r(\theta_m)
\ba_t^{\Herm}(\theta_m)
\bx_F[n,l]
e^{-j2\pi n\Delta f\tau_m} \\
& \times 
e^{j2\pi lT_{\mathrm{sym}}v_m}
+\bz[n,l],
\quad
\forall n\in[N_s],~\forall l\in[L],
\end{aligned}
\end{equation}
where $\alpha_m$, $\theta_m$, $\tau_m$, and $v_m$ denote the reflection coefficient, azimuth angle, delay, and Doppler frequency of target $m$, respectively.
The vectors $\ba_t(\cdot)$ and $\ba_r(\cdot)$ denote the transmit and receive steering vectors, respectively.
In addition, $\bz[n,l]\sim\mathcal{CN}(\mathbf{0},\sigma_s^2\bI_{N_r})$ denotes the additive white Gaussian noise.

Let $\boldsymbol{\theta}=[\theta_1, \theta_2, \ldots,\theta_M]^{\T}$
collect the target angles and $\boldsymbol{\eta}\!\in\!\R^{4M}\!$
collect the nuisance parameters $\{\tau_m,v_m,\Re\{\alpha_m\},\Im\{\alpha_m\}\}_{m=1}^{M}$.
The CRB for estimating $\boldsymbol{\theta}$ is given by
\begin{equation}
\label{eq:angle_crb}
\mathrm{CRB}_{\boldsymbol{\theta}}
=
\trace\!\left[
\left(
\bF_{\boldsymbol{\theta}\boldsymbol{\theta}}
-
\bF_{\boldsymbol{\theta}\boldsymbol{\eta}}
\bF_{\boldsymbol{\eta}\boldsymbol{\eta}}^{-1}
\bF_{\boldsymbol{\theta}\boldsymbol{\eta}}^{\T}
\right)^{-1}
\right].
\end{equation}
Let $\bs[n,l]$ denote the noiseless component of
\eqref{eq:sensing_received_signal}.
The FIM blocks in \eqref{eq:angle_crb} are given by
\[
	\bF_{\boldsymbol{\alpha}\boldsymbol{\beta}} = \frac{2}{\sigma_s^2} \sum_{l=0}^{L-1}\sum_{n=0}^{N_s-1}\Re\left\{ \frac{\partial\bs^{\Herm}[n,l]}{\partial\boldsymbol{\alpha}} \frac{\partial\bs[n,l]}{\partial\boldsymbol{\beta}^T} \right\}, \forall \boldsymbol{\alpha}, \boldsymbol{\beta} \in \{\boldsymbol{\theta}, \boldsymbol{\eta}\}.
\]

Direct minimization of $\mathrm{CRB}_{\boldsymbol{\theta}}$ is complicated by its dependence on all target parameters and the matrix inversions required in \eqref{eq:angle_crb}.
We therefore adopt a tractable lower bound, $M^2/\trace(\bF_{\boldsymbol{\theta}\boldsymbol{\theta}})$, as a surrogate objective \cite{chehab2008geometrical}.
Minimizing this bound is equivalent to maximizing
$\trace(\bF_{\boldsymbol{\theta}\boldsymbol{\theta}})$,
which serves as our sensing metric and can be expressed in the following quadratic form:
\begin{equation}
\label{eq:trace_fim}
\trace(\bF_{\boldsymbol{\theta}\boldsymbol{\theta}})
=
\frac{2}{\sigma_s^2}
\sum_{l=0}^{L-1}
\sum_{n=0}^{N_s-1}
\bx_F^{\Herm}[n,l]\bQ\bx_F[n,l],
\end{equation}
where
\[
	\bQ=
\sum_{m=1}^{M}
|\alpha_m|^2
\bB^{\Herm}(\theta_m)
\bB(\theta_m),~ ~  \bB(\theta)
= \frac{\partial  (\ba_r(\theta)\ba_t^\Herm(\theta) )}{\partial \theta}.
\]
The matrix $\bQ$ depends  on the target angles and reflection powers, and is independent of the target delays, Doppler frequencies, and reflection phases.
For waveform design, we construct $\bQ$ using the estimates $\{\hat{\theta}_m,|\hat{\alpha}_m|^2\}_{m=1}^{M}$.
Such estimates can be obtained from an initial sensing or parameter-estimation stage \cite{ren2024fundamental}, or predicted based on previous sensing observations in target-tracking scenarios \cite{liu2020radar}.

\subsection{Problem Formulation}
\label{subsec:formulation}

We aim to maximize the sensing metric subject to the  CI constraints and the time-domain 1-bit transmit signal constraint.
The waveform design problem is formulated as
\begin{equation}
	\label{eq:total_problem}
	\begin{aligned}
		\max_{\{\bx_F[l], \bx_T[l]\}} &\quad  \frac{1}{2}\sum_{l=0}^{L-1} \sum_{n=0}^{N_s-1}  \bx_F^\Herm[n,l] \bQ \bx_F[n,l] \\
		\st \hspace{1.5em} & \quad \eqref{eq:DFT relation},\eqref{eq:1-bit constraint},~\text{and}~\eqref{eq:CI constraint}.
	\end{aligned}
\end{equation}
Problem \eqref{eq:total_problem} is separable across OFDM symbols and can therefore be decomposed into $L$ independent subproblems. Suppressing the OFDM-symbol index $l$, each subproblem can be written as
\begin{subequations}
	\label{eq:separate_problem}
	\begin{align}
		\min_{\bx_F, \bx_T} &\quad \label{eq:separate_problem_obj}  -\frac{1}{2}\sum_{n=0}^{N_s-1} \bx_F^\Herm[n] \bQ \bx_F[n] \\
		\st \hspace{0.3em} & \label{eq:separate_problem_DFT} \quad \bx_F = \bG \bx_T, \\
		& \label{eq:separate_problem_CI} \quad \Re\{\bC_{n} \bx_F[n]\} \geq \boldsymbol{\gamma}, \quad \forall n\in[N_s], \\
		& \label{eq:separate_problem_1bit} \quad \bx_T \in \mathcal{X}^{N_sN_t}.
	\end{align}
\end{subequations} 
In problem \eqref{eq:separate_problem}, the objective function
\eqref{eq:separate_problem_obj} and the CI constraints \eqref{eq:separate_problem_CI} are separable across subcarriers
in the frequency domain, whereas the 1-bit transmit signal constraint
\eqref{eq:separate_problem_1bit} is imposed elementwise on the
time-domain samples.
The frequency- and time-domain variables are coupled through the DFT relation \eqref{eq:separate_problem_DFT}.

\section{Algorithmic Design}
This frequency--time domain structure motivates a variable-splitting approach. 
We therefore develop an inexact ADMM algorithm that alternates between the frequency-domain variable $\bx_F$ and the time-domain variable $\bx_T$, while enforcing their consistency through the DFT relation \eqref{eq:separate_problem_DFT}. 
Specifically, define the partial augmented Lagrangian associated with \eqref{eq:separate_problem_DFT} as
\begin{equation*}
\begin{aligned}
\mathcal{L}_{\rho}(\bx_F,\bx_T,\boldsymbol{\lambda})
=&-\frac{1}{2}\sum_{n=0}^{N_s-1} \bx_F^\Herm[n] \bQ \bx_F[n]+\Re\{\boldsymbol{\lambda}^\Herm(\bx_F-\bG\bx_T)\}\\ 
&+\frac{\rho}{2}\norm{\bx_F-\bG\bx_T}^2,
\end{aligned}
\end{equation*}
where $\boldsymbol{\lambda}\in\C^{N_sN_t}$ is the multiplier associated with \eqref{eq:separate_problem_DFT} and $\rho>0$ is the penalty parameter. 
At iteration $\ell$, the ADMM updates take the following form:
\begin{align}
    & \label{eq:ADMM subproblem xT} \begin{aligned}
        \bx_T^{(\ell+1)} & \gets \argmin_{\bx_T \in \mathcal{X}^{N_sN_t}} ~~ \mathcal{L}_{\rho^{(\ell)}}(\bx_F^{(\ell)},\bx_T,\boldsymbol{\lambda}^{(\ell)}),\\
    & \hspace{3em} \st \hspace{1.8em} \Re\{\bD\bx_T\} \geq \bar{\boldsymbol{\gamma}}.\\
    \end{aligned} \\
    & \label{eq:ADMM subproblem xF} \begin{aligned}
        \bx_F^{(\ell+1)} & \gets \argmin_{\bx_F \in \C^{N_sN_t}} ~~ \mathcal{L}_{\rho^{(\ell)}}(\bx_F,\bx_T^{(\ell+1)},\boldsymbol{\lambda}^{(\ell)}), \\
        & \hspace{3em} \st \hspace{1.8em}  \Re\{\bC_{n} \bx_F[n]\} \geq \boldsymbol{\gamma}, \quad \forall n\in[N_s]. \\
    \end{aligned} \\
    & \label{eq:lambda-update} \boldsymbol{\lambda}^{(\ell+1)} \gets ~~\boldsymbol{\lambda}^{(\ell)}+\rho^{(\ell)}\left( \bx_F^{(\ell+1)}-\bG\bx_T^{(\ell+1)}\right).
\end{align}
Note that we impose the time-domain CI constraint \eqref{eq:time-domain CI} in problem~\eqref{eq:ADMM subproblem xT}.
This helps promote the feasibility of the intermediate iterates, and
its effectiveness will be demonstrated by comparison with a conventional ADMM scheme that imposes the CI constraints only in the
frequency domain.

We solve both subproblems \eqref{eq:ADMM subproblem xT} and \eqref{eq:ADMM subproblem xF} inexactly to reduce computational cost, as described in the following subsections.

\subsection{Time-Domain Update for $\bx_T$}

We exploit the facts that $\bG$ is unitary and that every
$\bx_T\in\mathcal{X}^{N_sN_t}$ satisfies $\|\bx_T\|^2=N_sP$.
Dropping the terms independent of $\bx_T$ in
problem~\eqref{eq:ADMM subproblem xT} yields
\begin{subequations}
\label{eq:xT-subproblem}
\begin{align}
\min_{\bx_T\in\mathcal{X}^{N_sN_t}}
&\quad
-\Re\left\{
\left[
\bG^{\Herm}
\left(
\boldsymbol{\lambda}^{(\ell)}
+\rho^{(\ell)}\bx_F^{(\ell)}
\right)
\right]^{\Herm}\bx_T
\right\}\\
\st \hspace{1em}
&\quad
\Re\{\bD\bx_T\}\geq\bar{\boldsymbol{\gamma}}.
\label{eq:xT CI}
\end{align}
\end{subequations}
This constrained integer linear programming problem does not admit a closed-form solution.
We dualize the CI constraints using a nonnegative multiplier
$\boldsymbol{\nu}\in\R_+^{2KN_s}$.
The resulting Lagrangian is
\[
\begin{aligned}
\mathcal{L}_{T,\ell}(\bx,\boldsymbol{\nu})
={}&\boldsymbol{\nu}^{\T}\bar{\boldsymbol{\gamma}}-\Re\left\{
\left[
\bG^{\Herm}
\left(
\boldsymbol{\lambda}^{(\ell)}
+\rho^{(\ell)}\bx_F^{(\ell)}
+\bC^{\Herm}\boldsymbol{\nu}
\right)
\right]^{\Herm}\bx
\right\}.
\end{aligned}
\]
For a fixed $\boldsymbol{\nu}$, minimization of $\mathcal{L}_{T,\ell}$ over the 1-bit alphabet separates across entries and a minimizer is given by
\[
\bx_{T,\ell}^{*}(\boldsymbol{\nu})
=
\mathcal{P}_{\mathcal{X}}\left(
\bG^{\Herm}
\left(
\boldsymbol{\lambda}^{(\ell)}
+\rho^{(\ell)}\bx_F^{(\ell)}
+\bC^{\Herm}\boldsymbol{\nu}
\right)
\right),
\]
where
$$\mathcal{P}_{\mathcal{X}}(\cdot)= \sqrt{\frac{P}{2N_t}}\left[ \sgn(\Re\{\cdot\})+j\sgn(\Im\{\cdot\}) \right]. $$
Substituting this minimizer yields the Lagrangian dual problem
\begin{equation}
\label{eq:xT subproblem dual}
\max_{\boldsymbol{\nu}\in\R_+^{2KN_s}}
\quad
g_{\ell}(\boldsymbol{\nu})
\triangleq
\mathcal{L}_{T,\ell}
\bigl(\bx_{T,\ell}^{*}(\boldsymbol{\nu}),\boldsymbol{\nu}\bigr).
\end{equation}
We maximize $g_{\ell}(\boldsymbol{\nu})$ using projected subgradient ascent.
A subgradient of $g_{\ell}(\cdot)$ at $\boldsymbol{\nu}$ is
$\bar{\boldsymbol{\gamma}}
-\Re\{\bD\bx_{T,\ell}^{*}(\boldsymbol{\nu})\}$.
With $i$ denoting the inner iteration index, the multiplier update is
\begin{equation}
\label{eq:xT-dual-update}
\boldsymbol{\nu}^{(\ell,i+1)}
=\Big[
\boldsymbol{\nu}^{(\ell,i)}
+\beta_i\Big(
\bar{\boldsymbol{\gamma}}-\Re\big\{
\bD\bx_{T,\ell}^{*}
(\boldsymbol{\nu}^{(\ell,i)})
\big\}
\Big)
\Big]_+,
\end{equation}
where $\beta_i>0$ is the step size and $[\cdot]_+$ denotes projection onto the nonnegative orthant.

The dual ascent procedure is implemented inexactly using at most $I_\ell$ inner iterations, where $I_\ell$ is a prescribed iteration limit.
Let~$\boldsymbol{\nu}^{(\ell)}$ denote the final dual iterate. The time-domain update is recovered by
\begin{equation}
\label{eq:xT-recover}
\bx_T^{(\ell+1)}=\bx_{T,\ell}^{*}
\bigl(\boldsymbol{\nu}^{(\ell)}\bigr).
\end{equation}

\subsection{Frequency-Domain Update for $\bx_F$}
Given $\bx_T^{(\ell+1)}$, the frequency-domain subproblem \eqref{eq:ADMM subproblem xF} separates into $N_s$ independent $N_t$-dimensional complex quadratic programs. For each $n\in[N_s]$, we solve
\begin{subequations}
\label{eq:xF-block-subproblem}
\begin{align}
\min_{\bx_F[n]}
&\quad \frac{1}{2}\bx_F^{\Herm}[n]\bA_\ell\bx_F[n]
+\Re\left\{\bq_{n,\ell}^{\Herm}
\bx_F[n]\right\} \\
\st
&\quad \Re\left\{\bC_{n}\bx_F[n]\right\}
\geq \boldsymbol{\gamma},
\end{align}
\end{subequations}
where $\bA_{\ell} = \rho^{(\ell)}\bI_{N_t}-\bQ,
\bq_{n,\ell}
=
\boldsymbol{\lambda}_{n}^{(\ell)}
-\rho^{(\ell)}\bG_{n}\bx_T^{(\ell+1)}$, $\boldsymbol{\lambda}_{n}^{(\ell)}$ is the $n$-th $N_t$-dimensional block of $\boldsymbol{\lambda}^{(\ell)}$, and $\bG_{n}$ is the $n$-th row block of $\bG$.
Let $\zeta_{\max}(\bQ)$ denote the largest eigenvalue of $\bQ$.
Choosing $\rho^{(\ell)} > \zeta_{\max}(\bQ)$ ensures that $\bA_\ell \succ \boldsymbol{0}$ and that problem~\eqref{eq:xF-block-subproblem} is strictly convex.

Since the number of users $K$ is typically much smaller than the number of antennas $N_t$ in MIMO systems, we consider the dual problem of \eqref{eq:xF-block-subproblem}:
\begin{equation}
\label{eq:xF-block-dual}
\max_{~\boldsymbol{\mu}_{n} \in \R_+^{2K} }
\quad
-\frac{1}{2}\boldsymbol{\mu}_{n}^{\T}
\bM_{n,\ell}\boldsymbol{\mu}_{n}
+\bd_{n,\ell}^{\T}
\boldsymbol{\mu}_{n},
\end{equation}
where $\bM_{n,\ell}=\Re\{\bC_n\bA_{\ell}^{-1}\bC_n^\Herm\}$ and $\bd_{n,\ell}=\Re\{\bC_n\bA_{\ell}^{-1}\bq_{n,\ell}\}+~\!\boldsymbol{\gamma}.$
We solve \eqref{eq:xF-block-dual} inexactly using the
accelerated projected-gradient (APG) method, terminating at an
$\epsilon_\ell$-stationary point \cite{nesterov2013gradient}, where the prescribed tolerances satisfy $\epsilon_\ell\rightarrow0$ as
$\ell$ increases.
Let $\boldsymbol{\mu}_n^{(\ell)}$ denote the final dual
iterate for subcarrier $n$.
The frequency-domain update is recovered as
\begin{equation}
\label{eq:xF-recovery}
\bx_F^{(\ell+1)}[n]
=
\bA_\ell^{-1}
\left(
\bC_n^{\Herm}\boldsymbol{\mu}_n^{(\ell)}
-\bq_{n,\ell}
\right),
\quad \forall n\in[N_s].
\end{equation}

\begin{algorithm}[t] 
\caption{Inexact ADMM for solving \eqref{eq:separate_problem}} \label{alg:ADMM} 
\begin{algorithmic}[1] 
\State \textbf{Input:} $\ell_{\max},\rho_{\max},\epsilon,\epsilon', \{I_{\ell}\},\{\xi_\ell\},\{\epsilon_\ell\},\{\beta_i\}$.
\State Initialize $\bx_F^{(0)}, \bx_T^{(0)}, \boldsymbol{\lambda}^{(0)}, \rho^{(0)}\in(\zeta_{\max}(\bQ),\rho_{\max}), \ell \gets 0$, $r_0\gets \|\bx_F^{(0)}-\bG\bx_T^{(0)}\|, \delta_0\gets \|[\bar{\boldsymbol{\gamma}}-\Re\{\bD\bx_T^{(0)}\}]_+\|$. 
\For{$\ell=0,1,\ldots,\ell_{\max}-1$}
\State $i \gets 0, \boldsymbol{\nu}^{(\ell,0)}\gets \mathbf{0}$
\Repeat
\State Update $\boldsymbol{\nu}^{(\ell,i+1)}$ via \eqref{eq:xT-dual-update}.
\State $i \gets i+1$.
\Until{$\|\boldsymbol{\nu}^{(\ell,i)} - \boldsymbol{\nu}^{(\ell,i-1)}\| < \epsilon'$ or $i \geq I_\ell$.}
\State Recover $\bx_T^{(\ell+1)}$ via \eqref{eq:xT-recover}.
\ForAll{$n\in[N_s]$}  
\State Solve \eqref{eq:xF-block-dual} by APG to an $\epsilon_\ell$-stationary point.
\State Recover $\bx_F^{(\ell+1)}[n]$ via \eqref{eq:xF-recovery}.
\EndFor \State Compute $r_{\ell+1}$ and $\delta_{\ell+1}$.
\State Update $\boldsymbol{\lambda}^{(\ell+1)}$ and $\rho^{(\ell+1)}$ via \eqref{eq:lambda-update} and \eqref{eq:rho-update}, respectively.
\If{$r_{\ell+1} < \epsilon$ and $\delta_{\ell+1}< \epsilon$}
\State \textbf{break.}
\EndIf
\EndFor
\State \textbf{Return:} $\bx_T^{(\ell+1)}, \bx_F^{(\ell+1)}$.
\end{algorithmic}
\end{algorithm}

\subsection{Implementation and Complexity Analysis}
Algorithm~\ref{alg:ADMM} summarizes the overall procedure, where  $r_\ell \triangleq \|\bx_F^{(\ell)}-\bG\bx_T^{(\ell)}\|$ and $\delta_{\ell} \triangleq \| [ \bar{\boldsymbol{\gamma}}-\Re\{ \bD\bx_T^{(\ell)}\} ]_+ \|$ denote the primal residual and time-domain CI constraint violation, respectively.
We adaptively update the penalty parameter using the following rule \cite{andreani2008augmented}:
\begin{equation}
\label{eq:rho-update}
\rho^{(\ell+1)} =
\begin{cases}
    \min\{2\rho^{(\ell)}, \rho_{\max}\},
        & \text{if } r_{\ell+1} > \xi_{\ell} r_{\ell}, \\
    \rho^{(\ell)},
        & \text{otherwise}.
\end{cases}
\end{equation}
where $\xi_{\ell} \in (0,1)$ and $\rho_{\max}>2\zeta_{\max}(\bQ)$ are prescribed parameters.

The computational complexity per ADMM iteration is as follows.  Computing \eqref{eq:xT-dual-update} using fast Fourier transforms (FFTs) costs $\bigO\left( I_{\ell}N_sN_t\left(\log N_s+K\right) \right).$
The number of inner iterations to obtain an $\epsilon_\ell$-stationary point of \eqref{eq:xF-block-dual} using the APG method is $\bigO\left(\sqrt{\kappa_\ell} \log(1/\epsilon_\ell)\right)$ \cite{nesterov2013gradient}, where $\kappa_\ell$ is a common upper bound on the condition numbers of $ \{\bM_{n,\ell}\}$. 
The overall complexity per ADMM iteration is $\bigO ( I_{\ell}N_sN_t\left(\log N_s+K\right) + \sqrt{\kappa_\ell} N_s K^2\log(1/\epsilon_{\ell}) + N_sN_tK(N_t+K))$.

\section{Numerical Results}

The simulation settings are as follows.  
The MIMO-OFDM DFRC system has $N_t=8$ transmit antennas and $N_r=8$ receive antennas and transmits $L=10$ OFDM symbols over $N_s=32$ subcarriers per frame.
The BS serves $K=2$ communication users and senses $M=3$ targets, whose azimuth angles and reflection powers are $(\theta_1,\theta_2,\theta_3)=(-20^\circ,0^\circ,35^\circ)$ and $(|\alpha_1|^2,|\alpha_2|^2,|\alpha_3|^2)=(0.2,0.1,0.5)$.
The transmit power is $P=20$ dBm and the communication and sensing noise powers are $\sigma_c^2=0$ dBm and $\sigma_s^2=-10$~dBm, respectively.
The communication channels are generated according to the Rayleigh fading model $\bh_{k,n}\overset{\text{i.i.d.}}{\sim}\mathcal{CN}(\mathbf{0},\mathbf{I}) $.
QPSK modulation is employed with an SEP requirement of $\epsilon_c=1\times 10^{-4}$.

\begin{figure}[t]
	\centering
	\subfigure{\includegraphics[height=1.65in]{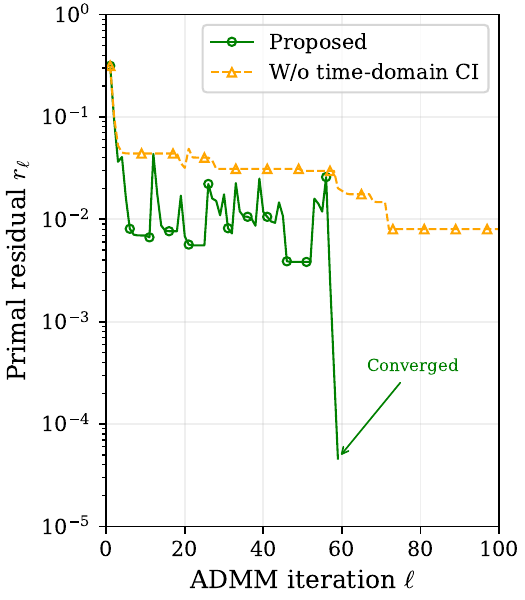}} \hspace{1em}
	\subfigure{\includegraphics[height=1.65in]{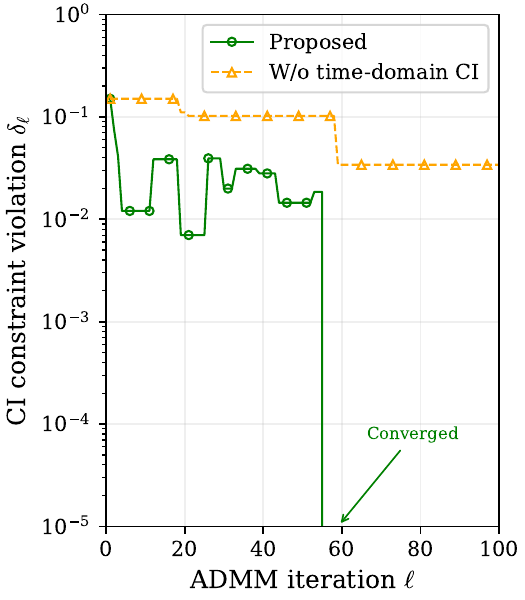}} 
	\caption{Convergence behavior of Algorithm~\ref{alg:ADMM} and its variant without the time-domain CI constraint \eqref{eq:time-domain CI}.}
    \label{fig:convergence}
\end{figure}

\subsection{Convergence Behavior}

First, we evaluate the convergence behavior of  Algorithm~\ref{alg:ADMM} in terms of the primal residual $r_{\ell}$ and the time-domain CI constraint violation~$\delta_{\ell}$.
As a benchmark, we also consider an ADMM variant that imposes only the frequency-domain CI constraints \eqref{eq:separate_problem_CI}.
The results are shown in Fig.~\ref{fig:convergence}.
It can be observed that the proposed ADMM algorithm converges rapidly, whereas the benchmark exhibits much slower convergence and larger violations of the  CI constraints.

\subsection{Sensing--Communication Performance}

\begin{figure}[t]
	\centering
	\subfigure{\includegraphics[height=1.65in]{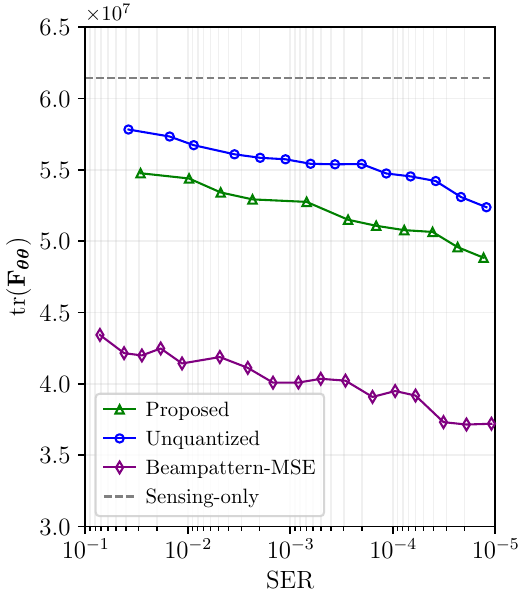}} \hspace{1em}
	\subfigure{\includegraphics[height=1.65in]{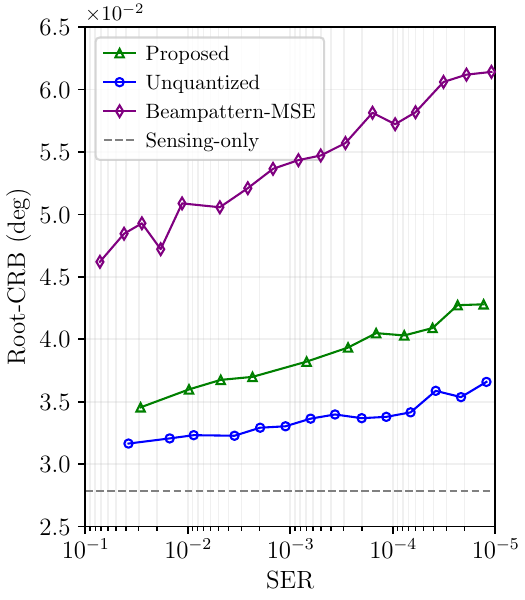}}
	\caption{Achieved sensing and communication tradeoff.}
    \label{fig:SC_tradeoff}
\end{figure}

Next, we compare the sensing--communication tradeoff obtained by the proposed method in terms of achieved symbol error rate (SER), the trace of the angular FIM and the square root of the CRB (in degrees) with those of the following three benchmarks: 
(1) \textbf{Unquantized}, which relaxes the 1-bit transmit signal constraint \eqref{eq:separate_problem_1bit} to $\bx_T \in \mathrm{conv}(\mathcal{X})^{N_sN_t}$;
(2) \textbf{Sensing-only}, which removes the communication CI constraints \eqref{eq:separate_problem_CI} from problem \eqref{eq:separate_problem};
(3) \textbf{Beampattern-MSE} \cite{wu2025quantized}, which minimizes the mean squared error (MSE) between the achieved and desired beampatterns, subject to the CI constraints \eqref{eq:separate_problem_CI} and the 1-bit transmit signal constraint \eqref{eq:separate_problem_1bit}. The desired beampattern is defined as
    \[
	    d(\theta)=\left\{
	        \begin{aligned}
	            1, &~\text{if~} \theta \in \cup_{m=1}^M (\theta_m-5^\circ, \theta_m+5^\circ),  \\
	            0, &~\text{otherwise}.
	        \end{aligned}
	    \right.
	\]
The results are shown in Fig.~\ref{fig:SC_tradeoff}, averaged over
100 independent system realizations. 
It is observed that the proposed design outperforms the Beampattern-MSE design across a range of achieved SERs, and closely approaches the  performance of the unquantized benchmark.

\newpage
\bibliographystyle{IEEEbib}
\bibliography{refs}

\end{document}